\documentclass[12pt]{article}

\input epsf

\begin{document}

\title {\bf The Relativistic Quantum Motions.}

\author{T.~Djama\thanks{E-mail:
{\tt djam\_touf@yahoo.fr}}}
\date{July 1, 2004}
\maketitle

\begin{abstract}
\noindent Using the relativistic quantum stationary
Hamilton-Jacobi equation within the framework of the equivalence
postulate, and grounding oneself on both relativistic and quantum
Lagrangians, we construct a Lagrangian of a relativistic quantum
system in one dimension and derive a third order equation of
motion representing a first integral of the relativistic quantum
Newton's law. Then, we plot the relativistic quantum trajectories
of a particle moving under the constant and the linear
potentials. We establish the existence of nodes and link them to
the de Broglie's wavelength.
\end{abstract}

\vskip\baselineskip



%
\vskip0.5\baselineskip \noindent {\bf 1\ Introduction}
\vskip0.5\baselineskip
%

Deriving quantum mechanics from an equivalence postulate, Faraggi
and Matone showed that the Schr\"odinger wave function must have
the form \cite{FM1,FM2,FM3,FM4}
\begin{equation}
\phi(x)=\left({\partial S_0 \over \partial x} \right)^{-{1 \over
2}}\; \left[\alpha\exp \left ({i \over \hbar   } S_0 \right)+
\beta\exp \left(-{i \over \hbar}S_0\right)\right]\; ,
\end{equation}
where $\alpha$ and $\beta$ are complex constants and $S_0$ is a
real function representing the quantum reduced action. They
established that the conjugate momentum given by
\begin{equation}
P={\partial S_0 \over \partial x}
\end{equation}
never vanishes for bound and unbound states making possible a
dynamical approach of the quantum motion of particles. This
conjugate momentum is always real even in classically forbidden
regions. They showed also, within the framework of differential
geometry \cite{FM1,FM2,FM3,FM4}, that the quantum stationary
Hamilton-Jacobi equation (QSHJE) which leads to the Shr\"odinger
equation is
\begin{equation}
{1 \over 2m_0}\left({\partial S_0 \over \partial x}\right)^2+
V(x)-E={\hbar^2 \over 4m_0} \left[{3 \over 2} \left({\partial S_0
\over \partial x}\right)^{-2} \left({\partial^2 S_0 \over
\partial x^2}\right)^2- \left({\partial S_0 \over \partial
x}\right)^{-1} \left({\partial^3 S_0 \over \partial x^3}\right)
\right]
\end{equation}
where $V(x)$ is the potential and $E$ the energy. The solution of
Eq. (3) investigated by Floyd \cite{Floyd1, Floyd2, Floyd3,
Floyd4} and Faraggi and Matone \cite{FM1,FM2,FM3,FM4} is given in
Ref. [9] as
\begin{equation}
S_0=\hbar \arctan\left(a{\theta \over \phi}+b\right)\; ,
\end{equation}
where $a$ and $b$ are real constants. $\theta$ and $\phi$ are two
real independent solutions of the Schr\"odinger equation. Taking
advantage on these results, Bouda and Djama have recently
introduced a quantum Lagrangian
\begin{equation}
L(x,\dot{x},\mu,\nu)={1\over 2} m {\dot{x}}^2 f(x,\mu,\nu)-V(x)
\; ,
\end{equation}
from which they derived the quantum law of motion \cite{BD1}.
They stated that the conjugate momentum of the non relativistic
and spinless particle is written as
\begin {equation}
\dot{x}\, {\partial S_0 \over \partial x}= {2(E-V)} \; .
\end {equation}
From this last equation, they derived the first integral of the
quantum Newton's law (FIQNL).
\begin {eqnarray}
(E-V)^4-{m{\dot{x}}^2 \over 2}(E-V)^3+{{\hbar}^2 \over 8}
{\left[{3 \over 2} {\left({\ddot{x} \over
\dot{x}}\right)}^2-{\dot{\ddot{x}} \over \dot{x}}
\right]} (E-V)^2\hskip15mm&& \nonumber\\
-{{\hbar}^2\over 8}{\left[{\dot{x}}^2 {d^2 V\over
dx^2}+{\ddot{x}}{dV \over dx}
 \right]}(E-V)-{3{\hbar}^2\over 16}{\left[\dot{x}
{dV \over dx}\right]^2}=0 \; ,
\end {eqnarray}
which goes at the classical limit $(\hbar \to 0)$ to the
classical conservation equation
\begin {equation}
{m\; \dot{x}^2 \over 2}+V(x)=E \; .
\end {equation}
Bouda and Djama have also plotted some trajectories of the
particle for several potentials \cite{BD2} .

The construction of the Lagrangian (5) and the establishment of
the fundamental dynamical equations (6) and (7) are important
steps to build a deterministic theory which restores the
existence of trajectories \cite{BD1,BD2,BD3} . Nevertheless, such
a formalism cannot approach both relativistic velocities cases,
and more than one dimension motions. The aim of this paper is to
generalize the dynamical formalism that we have recalled above
\cite{BD1,BD2,BD3} into the one dimensional relativistic
velocities cases. In this purpose let us recall the finding of
Faraggi, Matone and Bertoldi concerning the relativistic quantum
systems. They stated that the relativistic quantum wave function
is given by Eq. (1), where $S_0$ defines the relativistic quantum
reduced action, and wrote the relativistic quantum stationary
Hamilton-Jacobi equation (RQSHJ) as \cite{FM3,FM4}
\begin{eqnarray}
{1 \over 2m_0}\left({\partial S_0 \over \partial x}\right)^2-
{\hbar^2 \over 4m_0}\left[{3 \over 2} \left({\partial S_0 \over
\partial x}\right)^{-2} \left({\partial^2 S_0 \over \partial
x^2}\right)^2-\right.
\hskip35mm&& \nonumber\\
\left.\left({\partial S_0 \over \partial x}\right)^{-1}
\left({\partial^3 S_0 \over \partial x^3}\right) \right]+ {1
\over 2m_0c^2}\left[m_0^2c^4 -(E-V)^2\right]=0\; ,
\end{eqnarray}
where $V$ is the potential, $E$ is the total energy of the
particle of mass equal to $m_0$ at rest and $c$ is the light
velocity in vacuum. The solution of Eq. (9) can be expressed by
Eq. (4), where $\theta$ and $\phi$ represent now two real
independent solutions of the Klein-Gordon equation
\begin{equation}
-c^2 \hbar^2 {\partial^2 \phi \over \partial x^2}+
\left[m_0^2c^4-(E-V)^2\right]\phi(x)=0\; .
\end{equation}

Taking advantage on these results, we will introduce in the
following sections a relativistic quantum formalism with which we
study the dynamics of high energy particles. First, in Sec. 2, we
present a relativistic quantum Lagrangian from which we derive
the relativistic quantum law of motion. Then, in Sec. 3 we study
and plot the relativistic quantum trajectories (RQTs) of a
particle moving under the constant potential. In Sec. 4, we study
the linear potential case. Finally, in Sec. 5, we introduce our
definition of de Broglie's wavelength.

%
\vskip0.75\baselineskip \noindent {\bf 2\ \ The relativistic
quantum motion } \vskip0.5\baselineskip
%

Now, we consider a relativistic quantum system. As we have noticed
for the quantum systems \cite{BD1} , the relativistic quantum
reduced action $S_0$ expressed by Eq. (4) contains two constants
more than the usual constant $E$ appearing in the expression of
the classical reduced action. This suggests that the relativistic
quantum law of motion is a fourth order differential equation.
Then, as in the quantum case \cite{BD1}, we introduce in the
expression of the Lagrangian a function $f$ of $x$ depending on
the constants $a$ and $b$ playing the role of hidden parameters.
As the relativistic quantum Lagrangian must goes at the classical
limit $(\hbar \to 0)$ to the relativistic one, we postulate the
following form for the Lagrangian
\begin{equation}
L(x,\dot{x},a,b)=-m_0c^2 \sqrt{1-{\dot{x}^2 \over c^2}\;
f(x,a,b)}-V(x)\; ,
\end{equation}
in which the function $f(x,a,b)$ satisfies
\begin{equation}
 \lim_{\hbar \to 0} f(x,a,b)=1\; .
\end{equation}
Since $L$ depends only on the variables $x$, $\dot{x}$ and the
constants $a, b$, the conjugate momentum is given by
\begin{equation}
 P={\partial L \over \partial \dot{x}}=
{m_0\dot{x}f(x,a,b) \over \sqrt{1-(\dot{x}^2/c^2)\; f(x,a,b)}}\; .
\end{equation}
The Hamiltonian corresponding to the Lagrangian (11) is
\begin{equation}
 H=P\dot{x}-L\; .
\end{equation}
Replacing Eqs. (11) and (13) into Eq. (14), one obtains
\begin{equation}
H(x,P)=\sqrt{m_{0}^2c^4+{P^2c^2 \over f(x,a,b)}}+V(x)\; ,
\end{equation}
At the classical limit, we see clearly by using Eq. (12) that the
relativistic quantum momentum $P$ given in Eq. (13) reduces to the
relativistic one expressed as
\begin{equation}
 P={m_0\dot{x} \over
\sqrt{1-\dot{x}^2/c^2}}\; .
\end{equation}
Likewise, the Hamiltonian given in Eq. (15) reduces to his
relativistic form
\begin{equation}
H(x,P)=\sqrt{m_{0}^2c^4+P^2c^2 }+V(x)\; ,
\end{equation}
well known in special relativity. For the stationary cases the
Hamiltonian $H$ corresponds to the total energy $E$ of the
particle. Then, we can write Eq. (15) as
\begin{equation}
E=\sqrt{m_{0}^2c^4+{P^2c^2 \over f(x,a,b)}}+V(x)\; ,
\end{equation}
which reads
\begin{equation}
{1 \over 2m_0}\left({\partial S_0\over \partial x}\right)^2{1
\over f(x,a,b)}+{1 \over 2m_0c^2}\left[m_0^2c^4
-(E-V)^2\right]=0\; ,
\end{equation}
after taking into account of the Hamilton-Jacobi definition of the
conjugate momentum $P=\partial S_0 /\partial x$. By applying the
equivalence postulate, already introduced by Faraggi and Matone
in previous works \cite{FM1,FM2,FM3,FM4}, on \break Eq. (19) , we
obtain the RQSHJE given in Eq. (9). This result suggests strongly
that the introduction of a Lagrangian with the form (11) is
founded. Now, taking the expression of the function $f(x,a,b)$
from Eq. (13) into Eq. (19), we get
\begin{equation}
\dot{x}\, {\partial S_0 \over \partial x}=E-V(x) - {m_0^2 c^4
\over (E-V)}\; .
\end{equation}
The last equation represents the relativistic quantum law of
motion, so the relativistic quantum trajectories of a particle
moving under any potential $V(x)$ should be plotted using Eq.
(20). To proceed, one can deduce the expression of the conjugate
momentum from Eq. (4) and replace it into Eq. (20), then we can
integrate this equation to give the relativistic quantum time
equation $x(t)$. Note that at the classical limit $\hbar \to 0$,
Eq. (20) reduces to the relativistic conservation equation
\begin{equation}
E={m_0c^2 \over \sqrt{1-{\dot{x}^2/c^2}}}+V(x)\;
\end{equation}
since the conjugate momentum takes the form (16). Note also that
in the relativistic limit $(c \to \infty)$, the kinetic energy $
T=E-V-m_0c^2$
satisfies $ T\ll m_0c^2$, so , Eq. (20) reduces to
$$
\dot{x}\, {\partial S_0 \over \partial x}={2\; T}\; ,
$$
which is equivalent to Eq. (6) already established in Ref
\cite{BD1}.

\noindent Now,  deducing $\partial^2 S_0 /\partial x^2$ and
$\partial^3 S_0 /\partial x^3$ from Eq. (20) and replacing them
into the expression of the RQSHJE (Eq. (9)), we find the First
Integral of the Relativistic Quantum Newton's Law FIRQNL which
reads
\newpage
\begin{eqnarray}
[(E-V)^2-m_0^2c^4]^3\left[\left(1-{\dot{x}^2 \over
c^2}\right)-{m_0^2c^4\over
(E-V)^2}\right] \hskip30mm\nonumber\\
-{\hbar^2 \over 2}\left[{(E-V)^4-m_0^4c^8\over E-V
}\right]\left(\ddot{x}{dV \over dx}+ \dot{x}^2{d^2V \over
dx^2}\right)\hskip25mm\nonumber\\ +{\hbar^2 \over
2}[(E-V)^2-m_0^2c^4]^2 \left[{3 \over 2}\left({\ddot{x} \over
\dot{x}}\right)^2- {\dot{\ddot{x}} \over
\dot{x}}\right]\hskip20mm\nonumber\\
-{\hbar^2 \over 4}\left[4m_0^2c^4\left(1-{m_0^2c^4\over (E-V)^2
}\right)-\right.
\hskip15mm\nonumber\\
\left. 3\left(E-V+{m_0^2c^4\over E-V}
\right)^2\right]\left({\dot{x} {dV\over dx}}\right)^2=0
\end{eqnarray}
As we observe, Eq. (22) is a third order differential equation in
$x$ containing the first and the second derivatives of the
potential $V$ with respect to $x$. Then, its solution
$x(t,E,a,b,c)$ contains four integration constants which can be
determined by the initial conditions. It is clear that if we set
$\hbar=0$, Eq. (22) reduces to Eq.(21) representing the
relativistic conservation equation. Note also that, after taking
the relativistic limit $(c \to \infty)$ into Eq. (22), one gets
Eq. (7) representing the FIQNL.
\vskip0.75\baselineskip \noindent {\bf 3.\ \ Motion under the
constant potential } \vskip0.4\baselineskip
First, after using expression (4) of $S_0$, let us write the
dynamical equation (20) in following form
\begin{equation}
{dx \over dt}=\pm {1 \over \hbar a W}\left[E-V(x)-{m_0^2c^4 \over
E-V(x)} \right] \left[\phi_2^2+(a\phi_1+b\phi_2)^2\right] \; ,
\end{equation}
where $W$ represents the wronskian of the function $\phi_1$ and
$\phi_2$. The $\pm$ sign in Eq. (23) indicates that the motion
may be in either direction on the x axis \cite{BD2}. In the case
of a massive particle (not a photon) moving under a constant
potential $V=U_0$, we can distinct two kind of motions, the
classically permitted motion ($E-U_0\ll m_0c^2$) and the
classically forbidden one. First, we review the classically
permitted motion. Choosing the two solutions of the Klein-Gordon
equation (Eq. (10)) as
$$
\phi_1=\sin\left({\sqrt{(E-U_0)^2-m_0^2c^4}\;  \over \hbar
c}x\right)\, ,\, \phi_2=\cos\left({\sqrt{(E-U_0)^2-m_0^2c^4}\;
\over \hbar c}x\right), $$
%
%
\begin{figure}
\def\put(#1,#2)#3{\leavevmode\rlap{\hskip#1\unitlength\raise#2\unitlength\hbox{#3}}}
\centerline{ \vbox{\hsize=10.5cm\setlength{\unitlength}{1truecm}
\put(0,0){\epsfxsize=10cm \epsfbox{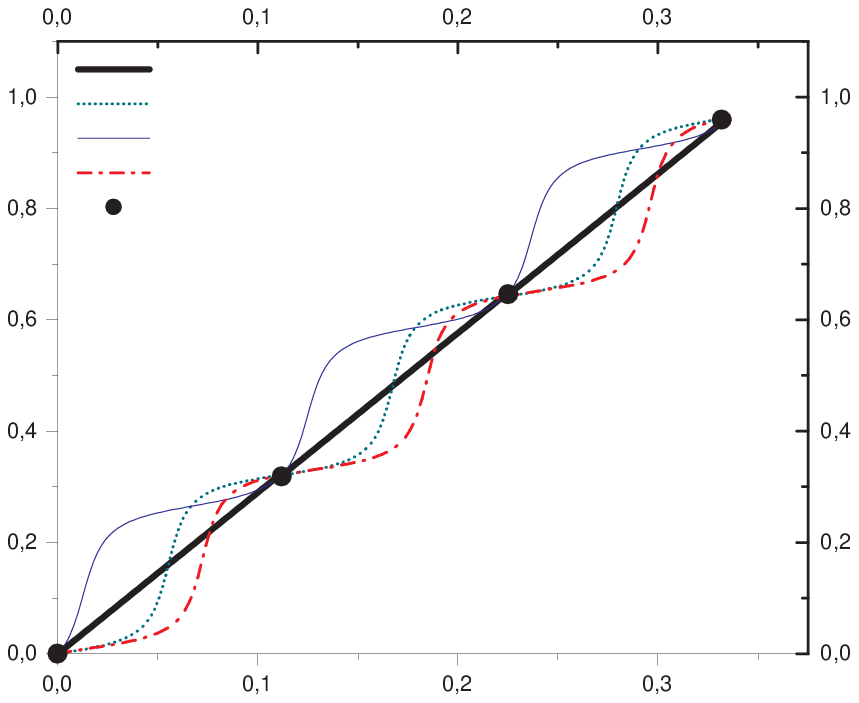}}
\put(8.5,0.6){$t$($\times 10^{-20}$ s)} \put(0.,7.95){$x$($\times
10^{-12}$ m)} \put(2.3,7.25) {\scriptsize purely relativistic
trajectory} \put(2.18,6.9) {\scriptsize $a=0.2,b=0$}
\put(2.08,6.55) {\scriptsize $a={4\over 3},b=-1.05$}
\put(2.0,6.2) {\scriptsize $a=0.25,b=8$} \put(1.88,5.8)
{\scriptsize Nodes} }} \centerline { \vbox { \hsize=10cm\noindent
\small Fig. 1:  Relativistic quantum trajectories for an electron
of energy $E=(2+U_0)$ Mev. For all the curves, we have chosen
$x(t=0)= 0$.} }
\end{figure}
and integrating the dynamical equation (23), we find
\begin{equation}
x(t)={\hbar c \over \sqrt{(E-U_0)^2-m_0^2c^4}}
\arctan{\left[{1\over a} \tan{\left({(E-U_0)^2-m_0^2c^4
\over\hbar (E-U_0)}\; t\right)}-{b\over a}\right]}+x_0 \; .
\end{equation}
This equation represents the time equation of RQTs. As we have
mentioned above, $x(t)$ contains four constants since the
fundamental equation of motion is a fourth order differential
equation. Because the arctangent function is defined on the
interval $\rbrack-\pi/2,\pi/2 \; \lbrack$, Eq. (24) shows that the
particle is contained between
$$
-{\hbar c  \over \sqrt{(E-U_0)^2-m_0^2c^4}} {\pi \over 2}+x_0
$$
and
$$
{\hbar c  \over \sqrt{(E-U_0)^2-m_0^2c^4}} {\pi \over 2}+x_0 \; .
$$
This is not possible. It is necessary to readjust the additive
integration constant $x_0$ after every interval of time in which
the tangent function goes from $-\infty$ to $+\infty$ in such a
way to guarantee the continuity of $x(t)$. In this purpose,
expression (24) must be rewritten as
\begin{eqnarray}
x(t)={\hbar c \over \sqrt{(E-U_0)^2-m_0^2c^4}}
\arctan{\left[{1\over a} \tan{\left({(E-U_0)^2-m_0^2c^4 \over\hbar
E-U_0}\; t\right)}-{b\over a}\right]}\hskip5mm\nonumber\\
+{\pi \hbar c \over \sqrt{(E-U_0)^2-m_0^2c^4}}n+x_0 \; .
\end{eqnarray}
with
$$
t\; \in \left[{\pi \hbar (E-U_0)  \over
(E-U_0)^2-m_0^2c^4}\left(n-{1 \over 2}\right) ; {\pi \hbar
(E-U_0)  \over (E-U_0)^2-m_0^2c^4}\left(n+{1 \over 2}\right)
\right]
$$
for every integer number $n$. The purely relativistic trajectory
is obtained for $a=1$ and $b=0$. Indeed, for these values, Eq.
(25) reduces to the relativistic relation
\begin {equation}
x(t)= {c \over E-U_0}\; \sqrt{(E-U_0)^2-m_0^2c^4}\; t+x_0  \; .
\end {equation}
In Fig. 1, we have plotted in $(t,x)$ plane for an electron of
energy $(2+U_0)Mev$ some trajectories for different values of $a$
and $b$. All these trajectories pass through nodes exactly as we
have seen for quantum trajectories of an electron moving under a
constant potential \cite{BD2}. These nodes correspond to the times
\begin {equation}
t_n={\pi \hbar (E-U_0)  \over (E-U_0)^2-m_0^2c^4}\left(n+{1 \over
2}\right)\; .
\end {equation}
The distance between two adjacent nodes is on time axis
\begin {equation}
\Delta t_n=t_{n+1}-t_n={\pi \hbar (E-U_0)  \over
(E-U_0)^2-m_0^2c^4}\; .
\end {equation}
and space axis
\begin {equation}
\Delta x_n=x_{n+1}-x_n={\pi \hbar c  \over
\sqrt{(E-U_0)^2-m_0^2c^4}}\; .
\end {equation}
These distances are both proportional to $\hbar$ meaning that at
the classical limit $(\hbar \to 0)$ the nodes becomes infinitely
near, and then, all quantum trajectories tend to the purely
relativistic one. As it is explained in Ref. \cite{BD2}, this is
the reason why in problems for which $\hbar$ can be disregarded,
relativistic quantum trajectories reduce to the purely
relativistic one \cite{BD1,BD2}. It is useful to note that at the
classical limit $(\hbar \to 0)$, not only the nodes becomes
infinitely near, but in addition the paths of all RQTs go to the
path of the purely relativistic one ($a=1, b=0$). Indeed, let us
consider an arbitrary point $P(x,t)$ from any RQT. Obviously,
this point is situated between two nodes. Now, if we take the
orthogonal projection $P_0$ of $P$ on the relativistic trajectory
($a=1$, $b=0$) and compute the distance $PP_0$ we get
\begin{equation}
PP_0=\sqrt{1+{(E-U_0)^2 \over c^2[(E-U_0)^2-m_0^2c^4]}}\,
|t_P-t_{P_0}|\; .
\end{equation}
\noindent Since Eq. (23) indicates that $\dot{x}$ is a monotonous
function, then, whatever the point $P$ we have $
|t_P-t_{P_0}|<t_{n+1}-t_{n}$ for every RQT \cite{BD2}. So, at the
classical limit $(\hbar \to 0)$, because $t_{n+1}-t_{n} \to 0$,
then $|t_P-t_{P_0}| \to 0$ and $PP_0 \to 0$. Thus, all RQTs go
when $(\hbar \to 0)$ to the purely relativistic one. This result
is in accordance with the fact that our dynamical equations (Eqs
(9), (20) and (22)) tend to the relativistic equations when
$\hbar \to 0$.

Now let us consider the classically forbidden motions $(E-U_0<0)$.
For this case the solution of Eq. (23) is
\begin{figure}
\def\put(#1,#2)#3{\leavevmode\rlap{\hskip#1\unitlength\raise#2\unitlength\hbox{#3}}}
\centerline{ \vbox{\hsize=10.5cm \setlength{\unitlength}{1truecm}
\put(0,0){\epsfxsize=10cm \epsfbox{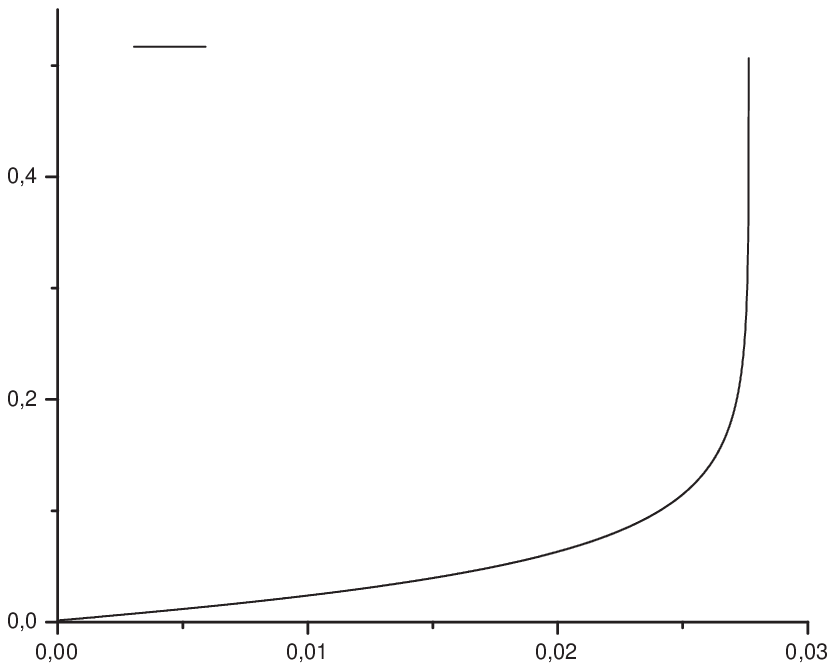}} \put(8.5,0.6) {$t$
($\times 10^{-20}$ s)} \put(0.2,7.75) {$x$ ($\times 10^{-12}$ m)}
\put(2.9,7.26) {\scriptsize $a=0.25,b=8$} }}
\centerline { \vbox
{ \hsize=12cm\noindent \small Fig. 2: Relativistic quantum
trajectories for an electron of energy $E=(U_0-2)$ Mev moving in
the classically forbidden regions. We have chosen $x(t=0)= 0$. In
the classically forbidden regions, the electron reach quickly an
infinite speed.} }
\end{figure}
%
%
\begin{equation}
x(t) = {\hbar \; c \over 2\sqrt{m_0^2c^4-(E-U_0)^2}}\ln
{\left\vert {1\over a} \tan \left({m_0^2c^4-(E-U_0)^2 \over\hbar
(E-U_0)}\; t\right)+{b \over a}\right\vert} + x_0 \; .
\end{equation}

We see clearly from Eq. (31) that at the finite time $$-(2n+1)\pi
\hbar {(E-U_0) \over4((E-U_0)^2-m_0^2c^4)}$$ the electron cross an
infinite distance and reach an infinite speed. This is in
accordance with standard quantum tunneling theories which predict
infinite velocities and finite reflexion times for tunneling
phenomena (Fletcher \cite{fle}, Hartman \cite{hart}). In Fig.
(2), we plotted  for an electron of energy $(U_0-2)Mev$ a RQT in
the classically forbidden regions with $a=0.25$ and $b=8$. This
figure shows clearly how the particle reach an infinite position
at a finite time. In particular, in the classically forbidden
regions there is no nodes.
\vskip0.5\baselineskip \noindent {\bf 4.\ \ The linear potential
case } \vskip0.5\baselineskip
Here, we investigate the motion of a massive particle (electron)
under a potential of the form
\begin {equation}
V(x) = gx  \; ,
\end {equation}
for which the Klein-Gordon equation takes the form
\begin{equation}
-c^2 \hbar^2 {\partial^2 \phi \over \partial x^2}+
\left[m_0^2c^4-(E-gx)^2\right]\phi(x)=0\; .
\end{equation}
To establish the RQTs for the linear potential case, we integrate
Eq. (23), where $\phi_1$ and $\phi_2$ are, now, two solutions of
Eq. (33).

In this paper, we do not present the analytic solutions of Eq.
(33), and in order to plot the RQTs, we approach the problem by
numeric methods. We, first, integrate numerically Eq. (33) to
obtain two independents solutions $\phi_1$ and $\phi_2$, then, we
plot the RQTs from Eq. (23). We opt in the two steps for the
Euler integration method. \noindent The RQTs for the linear
\begin{figure}
\def\put(#1,#2)#3{\leavevmode\rlap{\hskip#1\unitlength\raise#2\unitlength\hbox{#3}}}
\centerline{ \vbox{\hsize=10.5cm \setlength{\unitlength}{1truecm}
\put(0,0){\epsfxsize=10cm \epsfbox{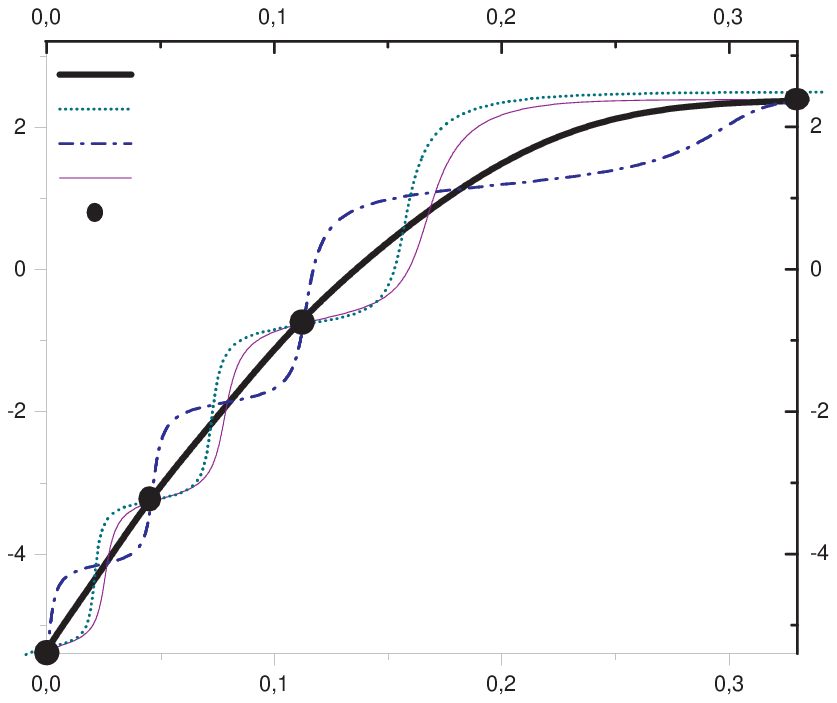}} \put(8.5,0.6) {$t$
($\times 10^{-20}$ s)} \put(0.,8.2) {$x$($\times 10^{-12}$ m)}
\put(2.3,7.35) {\scriptsize purely relativistic trajectory}
\put(2.18,7.0) {\scriptsize $a=4,b=2.5$} \put(2.08,6.62)
{\scriptsize $a=8,b=-3$} \put(2.0,6.3) {\scriptsize $a=5,b=2$}
\put(1.88,5.9) {\scriptsize Nodes} }} \centerline { \vbox {
\hsize=10cm\noindent \small Fig. 3: Relativistic quantum
trajectories for an electron of energy $E=2$ Mev moving under a
linear potential ($V(x)=gx$, $g=1$ Kg.m.s.). We have chosen
$x(t=0)= -5.4 \times 10^{-12}$ m.} }
\end{figure}
potential are presented in Fig. 3. We choose $E=2$ Mev, and $g=1$
Kg.m.s. In Ref. \cite{BD2}, we have chosen $g=10^{-9}$ Kg.m.s,
which is very small compared with $g=1$ Kg.m.s. We take this last
value for relativistic problem to render the quantity $gx$ of the
same order as $E$, so that the Klein-Gordon equation do not
reduces to the Schr\"odinger equation.

\noindent As we can notice from Fig. 3, the nodes are also
present for the linear potential case. The distance between two
adjacent nodes in Fig. 3 increases as the velocity decreases when
it approaches the turning point (point where the velocity
vanish). This note will be exposed in Sec. 5. Here we do not
investigate the classically forbidden regions. We would like to
stress that as for the linear potential in quantum cases, we
check that the positions of the nodes on $x$ axis are related to
the zeros  of the solution of the Schr\"odinger equation $\phi_2$
present in the denominator of the rapport in the expression (4)
of the reduced action $S_0$. This fact indicates that, the RQTs
likes the QTs and it is obvious that the QT are a limit of the
RQTs when $c \to \infty$.

\vskip0.5\baselineskip \noindent {\bf 5.\ \ De Broglie's
wavelength } \vskip0.5\baselineskip

One of the most important ideas that Secs. 3 and 4 bring is the
existence of nodes through which all RQTs pass, even the purely
relativistic one. In this section, we link the distance between
two adjacent nodes to the de Broglie's wavelength
\begin {equation}
\lambda = {h \over p}
\end {equation}
In Eq. (34), $p$ is the relativistic momentum. For a particle
moving under a constant potential
\begin {equation}
p = {\sqrt{(E-U_0)^2-m_0^2c^4} \over c} \; .
\end {equation}
By replacing Eq. (35) in Eq. (34), we get
\begin {equation}
\lambda = {hc \over \sqrt{(E-U_0)^2-m_0^2c^4}} \; .
\end {equation}
The distance between two nodes in the case of a constant potential
is
\begin {equation}
\Delta x_n={\pi \hbar c  \over \sqrt{(E-U_0)^2-m_0^2c^4}}\; .
\end {equation}
From Eqs. (36) and (37) we get
\begin {equation}
\Delta x_n= {\lambda \over 2} \; .
\end {equation}
Thus, the de Broglie's wavelength represents the double of the
distance between two adjacent nodes. As we have presented in Ref.
\cite{BD2}, we can generalize this definition for other
potentials. Indeed, if we compute the mean value of $\partial S_0
/ \partial x$ between two adjacent nodes, and taking into account
Eqs (25), (28) and (29) we find
\begin{eqnarray}
\left<{\partial S_0 \over \partial x }\right> \equiv {1 \over
\Delta x_n} \int_{x(t_n)}^{x(t_{n+1})} {\partial S_0 \over
\partial x } \; dx = {S_0(x(t_{n+1}))-S_0(x(t_n)) \over \Delta
x_n}\hskip-4mm\nonumber\\
 = {\sqrt{(E-U_0)^2-m_0^2c^4} \over c} \;
,
\end{eqnarray}
which is equal to $p$ (Eq. (35)). We propose to define a new
wavelength after substituting $p$ by
\begin {equation}
p = \left<{\partial S_0 \over \partial x }\right> \; .
\end {equation}
Then for any potential we can write, after using (2)
\begin {equation}
p = {\pi \hbar \over \Delta x} \; ,
\end {equation}
with $\Delta x$ is the distance between two adjacent nodes. If we
substitute (41) in (34) we find
\begin {equation}
\Delta x= {\lambda \over 2} \; .
\end {equation}
This relation links between the distance separating two adjacent
nodes and the de Broglie's wavelength whatever the potential
under which the particle moves.

\vskip\baselineskip {\bf Conclusions} To conclude, we would like
to stress that we exposed in this article an original approach of
the relativistic quantum mechanics. It is a generalization of the
one exposed in Refs. \cite{BD1,BD2,BD3}. So, we have derived the
fundamental relativistic quantum Newton's law expressed into Eqs.
(20) and (22). In addition, we have plotted the relativistic
quantum trajectories of particles moving under a constant
potential and a linear potential in both classically and
forbidden regions. For the classically permitted regions, we
established the existence of some nodes that we linked
successfully to the de Broglie's wavelength.

%
\vskip\baselineskip \noindent {\bf \ REFERENCES}
\vskip\baselineskip
%

\end{document}